\documentclass{article}
\usepackage{spconf,amsmath,graphicx,amsfonts}
\usepackage{multirow}
\usepackage{subfigure}
\usepackage{hyperref}
\usepackage[numbers,sort&compress]{natbib}

\title{Adversarial Attacks on Spoofing Countermeasures \\ of automatic speaker verification}
\name{Songxiang Liu$^1$, Haibin Wu$^2$, Hung-yi Lee$^2$, Helen Meng$^1$}

\address{
  $^1$Human-Computer Communications Laboratory, 
  The Chinese University of Hong Kong \\ 
  $^2$ Speech Processing and Machine Learning Laboratory, National Taiwan University \\
  }

\begin{document}
%

\maketitle
\begin{abstract}
High-performance spoofing countermeasure systems for automatic speaker verification (ASV) have been proposed in the ASVspoof 2019 challenge. However, the robustness of such systems under adversarial attacks has not been studied yet. 
In this paper, we investigate the vulnerability of spoofing countermeasures for ASV under both white-box and black-box adversarial attacks with the fast gradient sign method (FGSM) and the projected gradient descent (PGD) method. We implement high-performing countermeasure models in the ASVspoof 2019 challenge and conduct adversarial attacks on them. We compare performance of black-box attacks across spoofing countermeasure models with different network architectures and different amount of model parameters. The experimental results show that all implemented countermeasure models are vulnerable to FGSM and PGD attacks under the scenario of white-box attack. The more dangerous black-box attacks also prove to be effective by the experimental results. 
\end{abstract}
%


\section{Introduction}
\label{sec:intro}

Automatic speaker verification (ASV) aim to confirm that a given utterance is pronounced by a specified speaker. It is now a mature technology for biometric authentication \cite{snyder2018x,kanagasundaram2011vector,garcia2011analysis,senior2014improving,reynolds2000speaker,heigold2016end,lei2014novel,kenny2014deep}.
Modern speaker verification systems harness the combination of several modules to tackle the problem of ASV. In \cite{reynolds2000speaker}, for example, Gaussian mixture models (GMMs) are trained to model the acoustic features and likelihood ratio is used for scoring. 
Recently, ASV systems based on deep learning models require fewer concepts and heuristics compared to traditional speaker verification systems and have achieved considerable performance improvement. Heigold et al. \cite{heigold2016end} proposed an integrated model with end-to-end style which directly learns a mapping from a test utterance and a few reference utterances to a verification score, resulting in compact structure and sufficiently good performance. 
However, past research has shown that ASV systems are vulnerable to malicious attacks using spoofing and fake audios, such as synthetic, converted and replayed speech.

Anti-spoofing countermeasures for speaker verification systems arouse keen interests and several novel studies have been done \cite{evans2013spoofing,Todisco_2016,Valenti_2018,delgado2018asvspoof,wu2012study,khoury2014introducing,chen2015robust,lavrentyeva2017audio,chen2017resnet,qian2016deep,lai2019attentive}.
The ASVspoof 2019 challenge \cite{todisco2019asvspoof}, a community-led challenge, attracts more than 60 international industrial and academic teams to investigate spoofing countermeasures for ASV. Both the scenarios of logical access (LA) and physical access (PA) are taken into account. The LA scenario involves fake audios synthesized by modern text-to-speech synthesis (TTS) and voice conversion (VC) models. The PA scenario involves replayed audio recorded in reverberant environment under different acoustic and replay configurations. Several teams achieve excellent performance in detecting spoofing and reinforcing robustness of ASV systems under both LA and PA scenarios \cite{todisco2019asvspoof}. Yet according to \cite{szegedy2013intriguing,carlini2018audio}, machine learning models with impressive performance can be vulnerable to adversarial attacks \cite{goodfellow2014explaining}. Are the spoofing countermeasures for ASV in \cite{todisco2019asvspoof} robust enough to defend against adversarial examples?

An adversarial example $\tilde{x}$ is generated by the combination of a tiny perturbation and a normal instance $x$. It is very similar to the original normal instance $x$ and may even be visually or acoustically indistinguishable to human, but will lead the neural networks to incorrectly classify it as any target $t$ given a specific perturbation. Szegedy et al. \cite{szegedy2013intriguing} claim that well-trained neural network for image classification can succumb to adversarial attacks. 
The vulnerability of automatic speech recognition (ASR) neural network model under adversarial attacks is proved in \cite{carlini2018audio}, where an adversarial example can be transcribed as any phrase. 
However, to our best knowledge, the robustness of spoofing countermeasure systems for ASV under the adversarial attacks has not been studied yet. 
In this paper, we implement several high-performance spoofing countermeasure models in the ASVspoof 2019 challenge
and assess the reliability of these models under the attack of adversarial examples.

Adversarial attacks contains two main scenarios: white-box attack and black-box attack. White-box attacks are those where the adversary requires knowledge of the target model internals and adversarial examples are generated by an optimization strategy applied to the input space while fixing the model's parameters. 
Black-box attacks, such as \cite{papernot2017practical}, have no access to target model internals, only to its inputs and outputs. With the paired data acquired from available training data or by testing the online target model, a substitute for the target model can be trained. Then adversarial examples are easily crafted by the substitute and then used to attack the target model. The successful attack in the black-box scenarios, to some extent, guarantees the success of white-box attack because black-box attack requires less information and is more difficult than white-box attack. In this paper, for the sake of completeness, both white-box attacks and black box attacks are adopted to assess the reliability of spoofing countermeasure systems for ASV. There are a lot of adversarial attack approaches \cite{goodfellow2014explaining,carlini2017towards,moosavi2016deepfool,madry2017towards,su2019one}. In this paper, we adopt the fast gradient sign method (FGSM) \cite{goodfellow2014explaining} and the projected gradient descent (PGD) method \cite{madry2017towards}.

This paper is the first one investigating the vulnerability of spoofing countermeasures for ASV under both white-box and black-box adversarial attacks. 
We compare performance of black-box attacks across spoofing countermeasure models with different network architectures and different number of parameters. We implement countermeasure models which achieve comparable or even better anti-spoofing performance than some high-performance models in the ASVspoof 2019 challenge and we successfully attack them under both white-box and black-box attack scenarios. 
All our codes will be made open-source \footnote{Codes are available at \url{https://github.com/ano-demo/AdvAttacksASVspoof}}.

The paper is organized as follows. Section 2 provides the detailed description of two ASV spoofing countermeasure models. In section 3, we introduce the procedure of adversarial audio generation with two adversarial attack methods, i.e., the FGSM and the PGD method. In section 4, we report the experimental setups. The experiment results and discussion are given in section 5. Finally, we conclude this paper in section 6.

\section{ASV Spoofing Countermeasure Models}
\label{sec:asvCM}

Up to the submission time of this paper, only a few top-ranking systems in the ASVspoof 2019 challenge have accessible and complete model description. We choose two kinds of models, proposed by team T44 and team T45, to conduct adversarial attack experiments. According to the results of the ASVspoof 2019 challenge, the overall best performing single system for the LA scenario is proposed by team T45. The authors adopt an angular margin based softmax (A-softmax) \cite{liu2017sphereface} loss rather than traditional softmax with cross-entropy loss to train a Light CNN (LCNN) architecture \cite{wu2018light}. The system proposed by team T44 is ranked 3-rd and 14-th places for the PA and LA scenarios, respectively. Team T44 adopts a Squeeze-Excitation extended ResNet (SENet) as one of models in their submitted system. We provide brief description of these two kinds of models in the next two subsections.

\subsection{LCNN model and A-Softmax}

\begin{figure}[t!]
  \centering
  \centerline{\includegraphics[width=8.2cm]{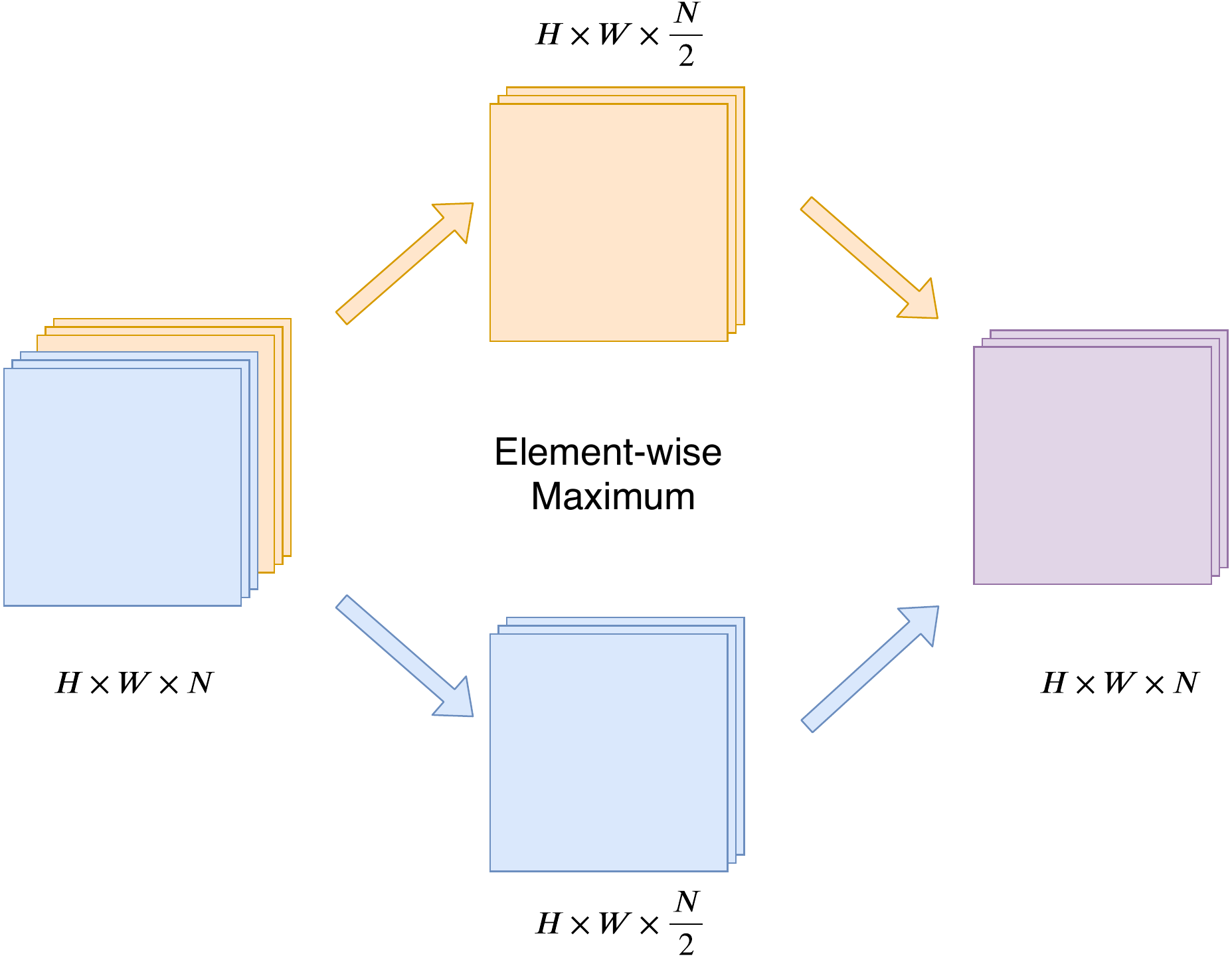}}
  \caption{The Max-Feature-Map (MFM) activation function for convolution layer.}
  \label{fig:MFM}
\end{figure}

The LCNN architecture is successfully adopted for replay attacks detection in \cite{lavrentyeva2017audio} and outperformed other proposed systems in ASVspoof 2017 challenge \cite{kinnunen2017asvspoof}. The LCNN architecture adopt the Max-Feature-Map (MFM) activation based on the Max-Out activation function. The MFM activation function for a convolutional layer is illustrated in Fig.~\ref{fig:MFM}, which is defined as
\begin{align}
\label{eq:1}
    &y_{i,j}^k = \max(x_{i,j}^k, x_{i,j}^{k+\frac{N}{2}}), \nonumber \\
    &\forall i = \overline{1, H}, j = \overline{1, W}, k = \overline{1, N/2},
\end{align}
where $x$ is the input feature map of size $H\times W \times N$, $y$ is the output feature map of size $H \times W \times \frac{N}{2}$.

To enhance the anti-spoofing performance of LCNN architecture, team T44 utilizes an A-Softmax loss to train the model. A-Softmax enables the model to learn angularly discriminative features. Geometrically, A-Softmax loss can be viewed as imposing discriminative constraints on a hypersphere manifold.
A-Softmax is represented as:
\begin{small}
\begin{equation}
\label{eq:2}
    L_{ang} \\
    = \frac{1}{N} \sum_{i} -\log(\frac{e^{||x_i||\cos(m\theta_{y_i,i})}}{e^{||x_i||\cos(m\theta_{y_i,i})} + \sum_{j\neq y_i}e^{||x_i||\cos(\theta_{y_j,i})}}),
\end{equation}\end{small}
where $N$ is the number of training samples, $\{x_i\}_{i=1}^{N}$ and their labels $\{y_i\}_{i=1}^{N}$ are training pairs, $\theta_{y_i,i}$ is the angle between $x_i$ and the corresponding column $y_i$ of weights $W$ in the fully connected classification layer, and $m$ is an integer that controls the size of an angular margin between classes.

\subsection{Squeeze-Excitation ResNet model}

\begin{figure}[t!]
  \centering
  \centerline{\includegraphics[width=8.2cm]{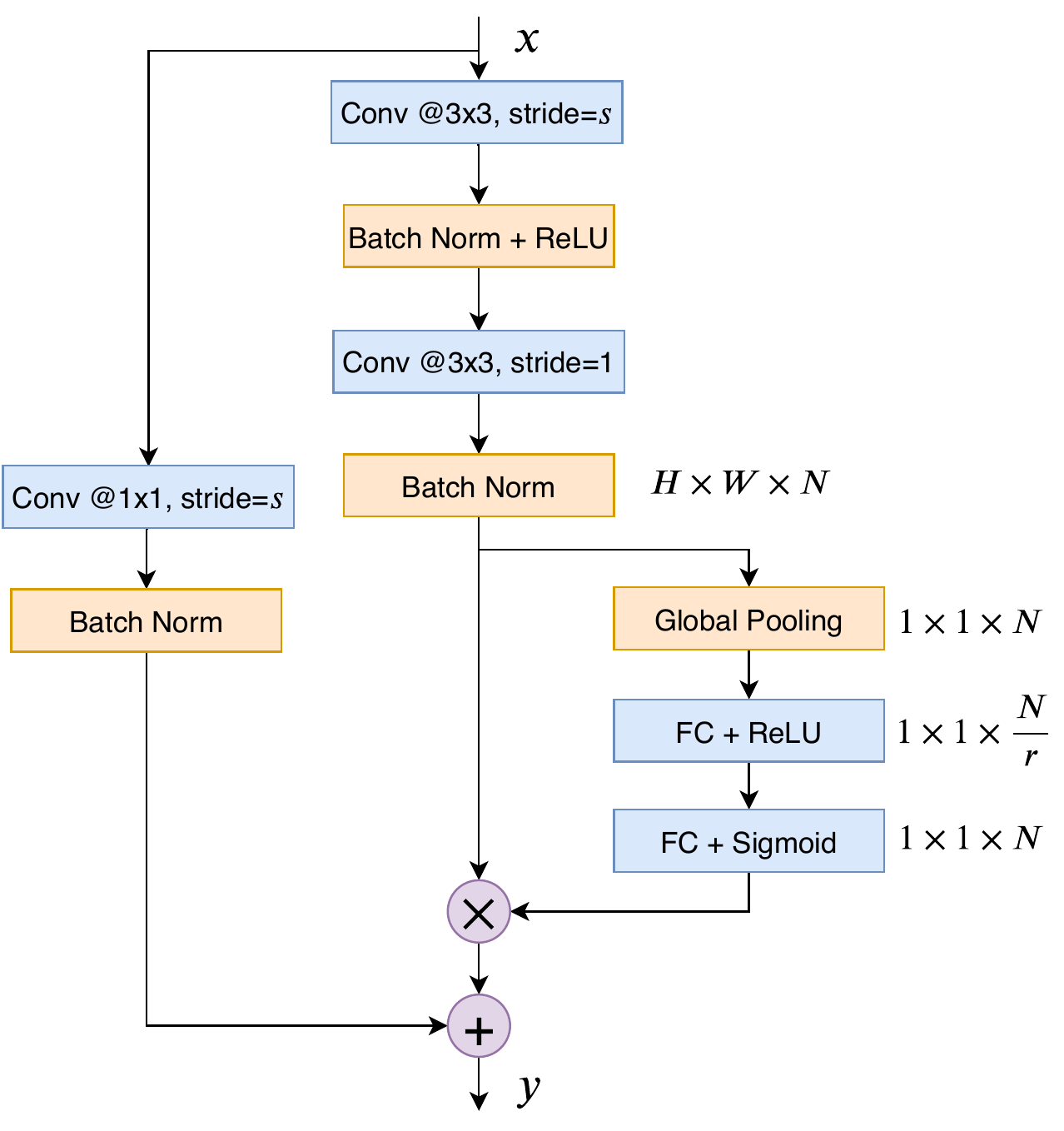}}
  \caption{The Squeeze-Excitation network (SENet) module, where $s$ is a customized variable specifying the stride and $r$ is the reduction factor. }
  \label{fig:senet}
\end{figure}

Squeeze-Excitation network (SENet) adaptively recalibrates channel-wise feature responses by explicitly modelling dependencies between channels, which has shown great merits in image classification task \cite{hu2018squeeze}. In this paper, we implemented a model with network architecture similar to that of SENet34 in \cite{lai2019assert}. The overall architecture of an SENet moduel is illustrated in Fig.~\ref{fig:senet}.

\section{Adversarial Audio Generation}
\label{sec:adv}
To execute an adversarial attack, we consider the model parameters $\theta$ as fixed and optimize over the input space. Specifically, in this paper, an adversarial example is a perturbed version of the input spectral feature $x$.
\begin{equation}
\label{eq:3}
    \tilde{x} = x + \delta, 
\end{equation}
where $\delta$ is small enough such that the reconstructed speech signal of the perturbed version $\tilde{x}$ is perceptually indistinguishable from the original signal $x$ by humans, but causes the network to make incorrect decision. Searching for a suitable $\delta$ can be formulated as solving the following optimization problem:
\begin{equation}
\label{eq:4}
    \max_{\delta \in \Delta}{\text{Loss}(\theta, x+\delta, y_x)},
\end{equation}
where $\text{Loss}(\cdot)$ is the loss function, $y_x$ is the label of $x$ and $\Delta$ is a set of allowed perturbation that formalizes the manipulative power of the adversary. In this paper, $\Delta$ is a small $l_{\infty}$-norm ball, that is, $\Delta = \{\delta| \, ||\delta||_{\infty}\leq \epsilon\}, \, \epsilon \geq 0 \in \mathbb{R}$. 

To solve the optimization problem (\ref{eq:4}), in this paper, we adopt the fast gradient sign method (FGSM) \cite{goodfellow2014explaining} and the projected gradient descent (PGD) method \cite{madry2017towards}.

\subsection{Fast gradient sign method}
The first adversarial attack method, shorted as the FGSM method, consists of taking a single step along the direction of the gradient, i.e.,
\begin{equation}
    \label{eq:5}
    \delta = \epsilon \cdot \text{sign}(\nabla_x \text{Loss}(\theta, x, y_{x})),
\end{equation}
where the $\text{sign}(\cdot)$ function simply takes the sign of the gradient $\nabla_x \text{Loss}(\theta, x, y_x)$. Therefore, given an utterance $x$, the adversarial spectral feature can be simply computed as
$\tilde{x} = x + \epsilon \cdot \text{sign}(\nabla_x \text{Loss}(\theta, x, y_{x}))$. While the FGSM benefits from being the simplest adversarial attack method, it is often relatively inefficient at solving the maximization problem (\ref{eq:4}). 

\subsection{Projected gradient descent method}

Unlike the FGSM, which is a single-step method, the PGD method is an iterative method. Starting from the original input $x_0 = x$, the input is iteratively updated as follows:
\begin{align}
    \label{eq:6}
    x_{k+1} = \text{clip}(x_k + \alpha \cdot \text{sign}(\nabla_{x_k} \text{Loss}(\theta, x_k, y_{x_k}))), \nonumber
    \\ \text{for} \, k = 0,...,K-1, 
\end{align}
where $\alpha$ is the step size, $K$ is the number of iterations and the $\text{clip}(\cdot)$ function applies element-wise clipping such that $||x_k - x||_{\infty}\leq \epsilon, \, \epsilon \geq 0 \in \mathbb{R}$. We take $x_K$ as the final perturbed spectral feature. Intuitively, the PGD method can be thought of as iteratively applying small-step FGSM, but forcing the perturbed input stay within the admissible set $\Delta$ at every step. The PGD method allows for more effective attacks but is naturally more computationally expensive than the FGSM. 

The performance of PGD is still limited by the possibility of sticking at local optima of the loss function. To mitigate this problem, random restarts is incorporated into the PGD method \cite{madry2017towards}. The PGD method with random restarts will be executed multiple runs. The initial location of the adversarial example is randomly selected within the admissible perturbation set $\Delta$ and the PGD method will be executed a certain number of times in one run. The final adversarial example is the one resulting in maximum loss.

\section{Experimental setups}

This paper uses the ASVspoof 2019 dataset, which encompasses partitions for the assessment of LA and PA scenarios. In this paper, we only utilize the LA partition, which are themselves partitioned into training, development and evaluation sets. 
We use raw log power magnitude spectrum computed from the signal as acoustic features. Following \cite{lavrentyeva2019stc}, FFT spectrum is extracted with the Blackman window having size of 1724 and step-size of 0.0081s. Only the first 600 frames for each utterance are used as input for all trained models. No additional preprocessing techniques such as voice activity detection (VAD), pre-emphasis or dereverberation are adopted.

The adversarial attacks are conducted as the following: We train a spoofing countermeasure model using the training set. 
Hyper-parameters of the countermeasure model are tuned using the development set. We evaluate the anti-spoofing performance and generate adversarial examples using the evaluation set. 
When generating adversarial examples, we add adversarial perturbation into the acoustic feature vectors using the FGSM or the PGD method introduced in Section \ref{sec:adv}. 
The perturbed acoustic features are reconstructed into waveform to attack the well trained countermeasure model.

\subsection{Details of countermeasure model implementation}

Three countermeasure models are trained, which we term as LCNN-big, LCNN-small and SENet12. LCNN-big has the same network architecture as in \cite{lavrentyeva2019stc}. LCNN-small has similar network architecture to LCNN-big, but with less parameters. SENet12 has similar network architecture as in \cite{lai2019assert} but with less parameters. The number of parameters of these three models are shown as in Table~\ref{tab:num_papa}. LCNN-big model has larger model capacity than LCNN-small and SENet12 model in terms of number of model parameters, while LCNN-small model and SENet12 model have equal model capacity. The detailed network architecture of LCNN-small and SENet12 model are shown in Table~\ref{tab:lcnn_small} and Table~\ref{tab:senet12}, respectively.

LCNN-big and LCNN-small model are trained using the A-Softmax loss function, while SENet12 is trained using the original softmax and cross-entropy loss. The constant $m$ in Eq.(\ref{eq:2}) is set to 4. To mitigate the overfitting issue, a dropout rate of 0.75 is used when training LCNN models. We found that adding a relatively large L2 regularization is helpful to mitigate the overfitting issue. We set the weight decay rate at 0.001 when training all these three models. We use Adam optimizer with a constant learning rate of 0.001 to update the model parameters in all training cases, where $\beta_1=0.9$ and $\beta_2=0.999$. 

During training stage, we applied early stopping according to the classification accuracy on the development set. During inference stage, we took the cosine similarity between the input and the weight vector in the last FC layer corresponding to the bonafide class as the score of the utterance.

\begin{table}[t!]
    \centering
    \caption{Number of parameters of LCNN-Big, LCNN-Small and SENet12 model.}
    \vspace{+0.5cm}
    \resizebox{\columnwidth}{!}{
    \begin{tabular}{|c|c|c|c|}
    \hline
    Model                 &  LCNN-big   & LCNN-small  & SENet12  \\
    \hline
    Num. of parametes     &  10M        & 0.51M       & 0.48M    \\
    \hline
    \end{tabular}}
    \label{tab:num_papa}
\end{table}

\begin{table}[t!]
    \centering
    \caption{LCNN-small network architecture.}
    \vspace{0.5cm}
    \begin{tabular}{c|c|c}
    \hline
    Type &  Filter / Stride & Output \\
    \hline 
    Conv\_1     & $5\times5$ / $1\times1$    & $863\times600\times16$  \\
    MFM\_2      & $-$            & $863\times600\times8$          \\
    \hline
    MaxPool\_3  & $2\times2 / 2\times2$    & $431\times300\times8$  \\
    \hline
    Conv\_4     & $1\times1 / 1\times1$    & $431\times300\times16$  \\
    MFM\_5      & $-$                      & $431\times300\times8$          \\
    BatchNorm\_6 & $-$                     & $431\times300\times8$            \\
    Conv\_7     & $3\times3 / 1\times1$    & $431\times300\times24$   \\
    MFM\_8      & $-$                      & $431\times300\times12$          \\
    \hline
    MaxPool\_9  & $2\times2 / 2\times2$     & $215\times150\times12$ \\
    BatchNorm\_10 & $-$                     & $215\times150\times12$            \\
    \hline
    Conv\_11     & $1\times1 / 1\times1$    & $215\times150\times24$  \\
    MFM\_12      & $-$                      & $215\times150\times12$          \\
    BatchNorm\_13 & $-$                     & $215\times150\times12$            \\
    Conv\_14     & $3\times3 / 1\times1$    & $215\times150\times24$   \\
    MFM\_15      & $-$                      & $215\times150\times12$          \\
    \hline
    MaxPool\_16  & $2\times2 / 2\times2$    & $107\times75\times12$ \\
    \hline
    Conv\_17     & $1\times1 / 1\times1$    & $107\times75\times24$  \\
    MFM\_18      & $-$                      & $107\times75\times12$          \\
    BatchNorm\_19 & $-$                     & $107\times75\times12$            \\
    Conv\_20     & $3\times3 / 1\times1$    & $107\times75\times8$  \\
    MFM\_21      & $-$                      & $107\times75\times4$          \\
    BatchNorm\_22 & $-$                     & $107\times75\times4$            \\
    Conv\_23     & $1\times1 / 1\times1$    & $107\times75\times8$  \\
    MFM\_24      & $-$                      & $107\times75\times4$          \\
    BatchNorm\_25 & $-$                     & $107\times75\times4$            \\
    Conv\_26     & $3\times3 / 1\times1$    & $107\times75\times8$  \\
    MFM\_27      & $-$                      & $107\times75\times4$          \\
    \hline
    MaxPool\_28  & $2\times2 / 2\times2$    & $53\times37\times4$ \\
    \hline
    FC\_29        & $-$            & 64  \\
    MFM\_30       & $-$            & 32          \\
    BatchNorm\_31 & $-$            & 32            \\
    \hline
    FC\_32        & $-$    & 2  \\
    \hline
    \end{tabular}
    \label{tab:lcnn_small}
\end{table}

\begin{table}[t!]
    \centering
    \caption{SENet12 network architecture.}
    \vspace{0.5cm}
    \begin{tabular}{c|c|c}
    \hline
   Type &  Filter / Stride & Output \\
    \hline 
    Conv     & $7\times7$ / $2\times2$    & $431\times300\times16$  \\
    BatchNorm & $-$                       & $431\times300\times16$  \\
    ReLU      & $-$                       & $431\times300\times16$  \\
    MaxPool  & $3\times3 / 2\times2$      & $215\times150\times16$  \\
    \hline
    SEResNet Module$\times1$ & $-$        & $215\times150\times16$  \\
    \hline
    SEResNet Module$\times2$ & $-$        & $107\times75\times32$  \\
    \hline
    SEResNet Module$\times3$ & $-$        & $53\times37\times64$  \\
    \hline
    SEResNet Module$\times1$ & $-$        & $26\times18\times128$  \\
    \hline
    Global AvgPool   & $-$                & 128  \\
    \hline
    FC               & $-$                & 2    \\
    \hline
    \end{tabular}
    \label{tab:senet12}
\end{table}

\subsection{Adversarial attacks}

We apply both white-box and balck-box adversarial attacks on the trained countermeasure models with the FGSM and the PGD method. 
We investigate adversarial attacks under various levels of manipulative power of the adversary on the audios. In both the FGSM and PGD attack settings, $\epsilon$ in Eq.(\ref{eq:5}) and Eq.(\ref{eq:6}) is chosen from the set of \{0.1, 1, 5\}. 
To make the level of manipulative power of the PGD attack consistent with the FGSM attack, we make the step-size $\alpha$ and $\epsilon$ in the PGD attack scenario satisfy the relationship of
\begin{equation}
    \epsilon=\text{number of iterations} \times \alpha.
\end{equation}
For example, if $\epsilon=1$ and number of iterations is 10, then $\alpha$ is set to be 0.1.
The number of random restarts is 5 in all experiments. 

\subsection{XAB listening test}
\label{subsec:XAB}

To achieve a valid adversarial attack, it is important to make the adversarial audio examples sound indistinguishable from the original audio signals by human ears. We conduct an XAB listensing test, which is a standard way to assess the detectable differences between two choices of sensory stimuli. The adversarial audio signals are generated from the LCNN-big using the PGD method with $\epsilon=5$. Each of the adversarial audio signal is reconstructed from the perturbed log power magnitude spectrum and the phase spectrum of its corresponding original audio signal. We presented to listeners 50 randomly chosen adversarial-original audio pairs (i.e., A and B), from each of which we randomly choose one as the reference audio (i.e., X). Five listeners take part in the XAB listening test, where they are asked to choose from A and B one audio which sounds more like the reference audio X.

\section{Results and discussion}

\subsection{Countermeasure performance}

Anti-spoofing performance of the three countermeasure models, LCNN-Big, LCNN-Small and SENet12, is evaluated via the minimum normalized tandem detection cost function (t-DCF) and the equal error rate (EER), as shown in Table~\ref{tab:cmresult}. Note that the LCNN-Big model achieves comparable performance with that reported in \cite{lavrentyeva2019stc}, and the SENet12 model has even better performance than the best performing single model reported in \cite{lai2019assert}.

\begin{figure}[t]
\centering
\subfigure[Reference.]{
\begin{minipage}[t]{0.33\linewidth}
\centering
\includegraphics[width=2.85cm]{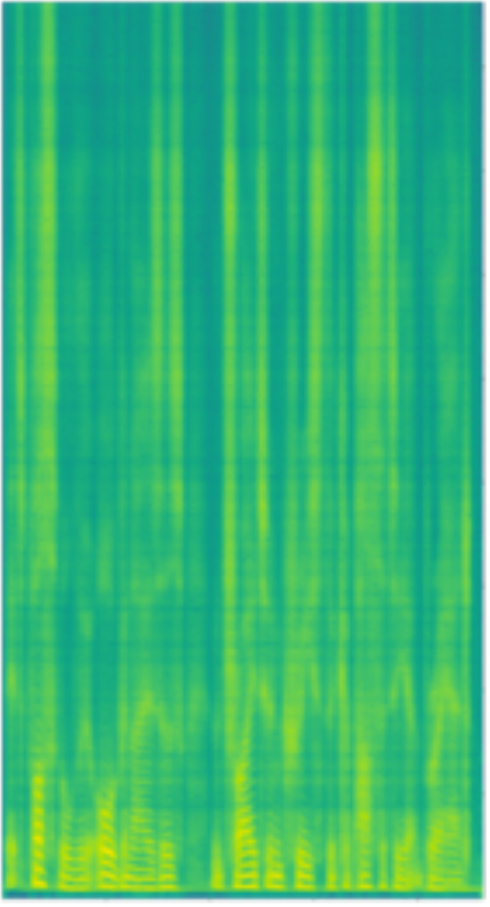}
\end{minipage}%
}%
\subfigure[Adv. example.]{
\begin{minipage}[t]{0.33\linewidth}
\centering
\includegraphics[width=2.85cm]{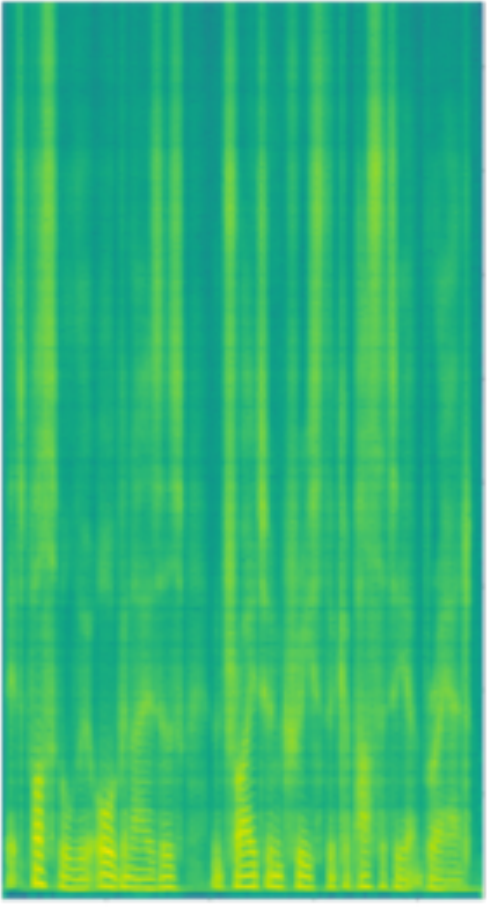}
\end{minipage}%
}%
\subfigure[Adv. perturbation.]{
\begin{minipage}[t]{0.33\linewidth}
\centering
\includegraphics[width=2.85cm]{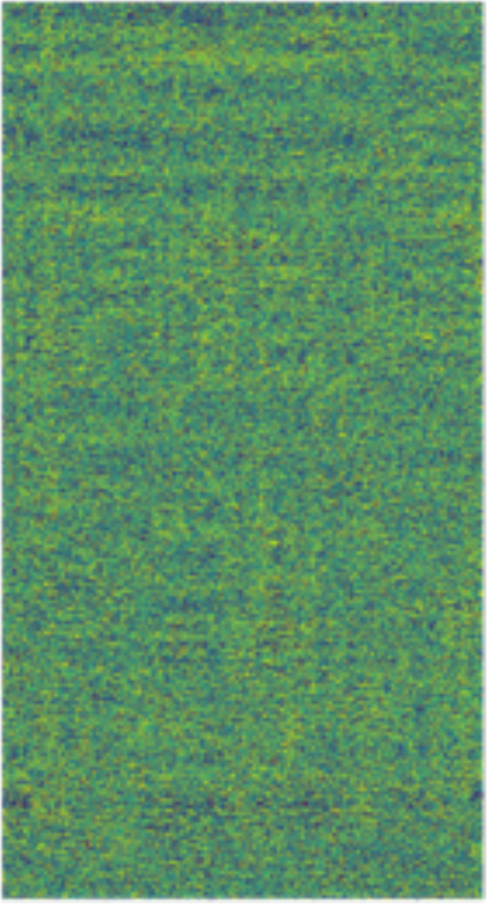}
\end{minipage}
}%
\centering
\caption{Sub-figures (a)-(c) are spectrograms of original audio, adversarial audio and adversarial perturbation of the utterance LA\_E\_1001227. The attack is conducted to the LCNN-big model using the PGD method, where $\epsilon=5$, number of iterations is 10 and number of random restarts is 5.}
\label{fig:spec}
\end{figure}

\begin{figure*}[t]
\centering
\subfigure[LCNN-big and LCNN-small.]{
\begin{minipage}[t]{0.33\linewidth}
\centering
\includegraphics[width=5cm]{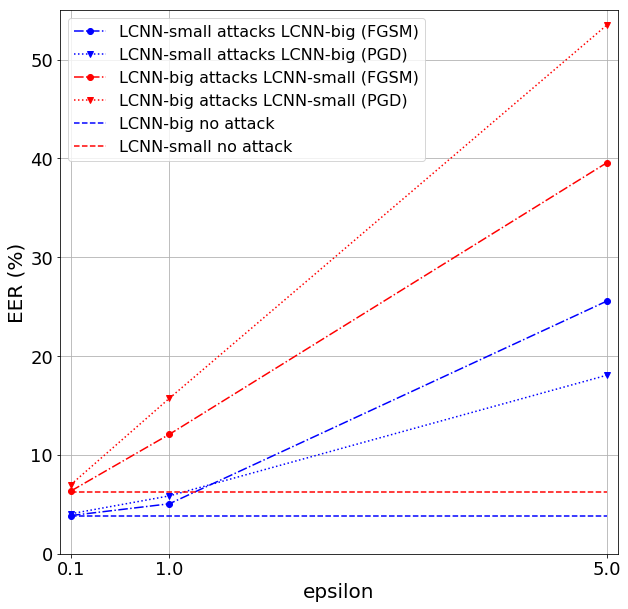}
\label{fig:Black-box attack.a}
\end{minipage}%
}%
\subfigure[LCNN-big and SENet12.]{
\begin{minipage}[t]{0.33\linewidth}
\centering
\includegraphics[width=5cm]{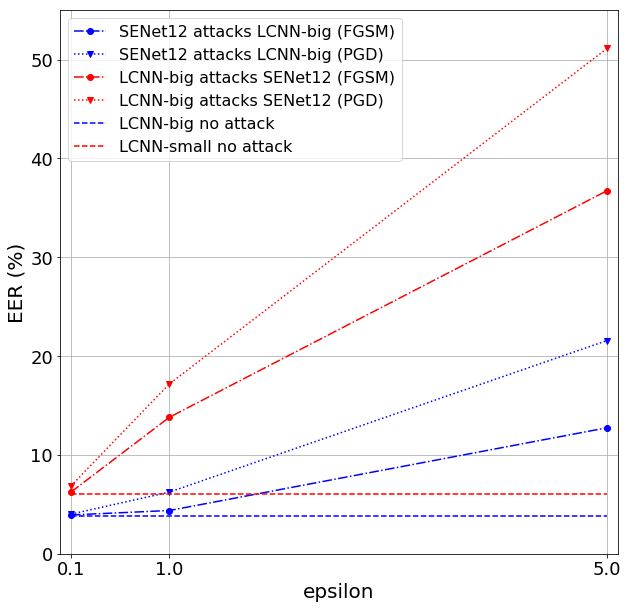}
\label{fig:Black-box attack.b}
\end{minipage}%
}%
\subfigure[LCNN-small and SENet12.]{
\begin{minipage}[t]{0.33\linewidth}
\centering
\includegraphics[width=5cm]{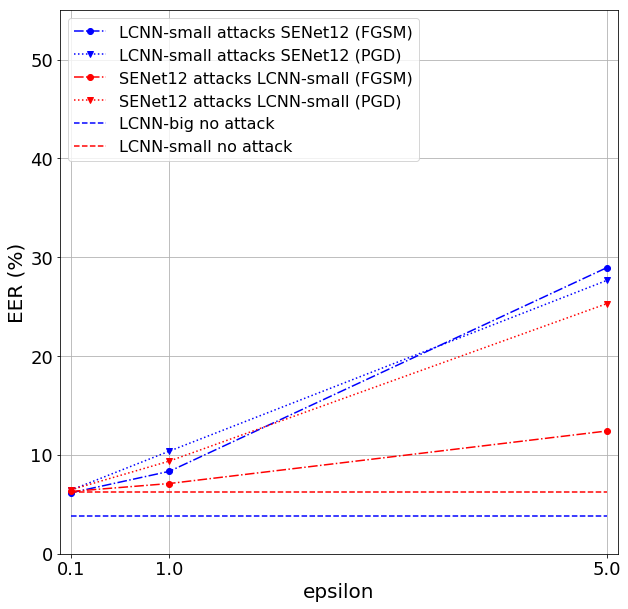}
\label{fig:Black-box attack.c}
\end{minipage}
}%
\centering
\caption{Black-box attack performance of the FGSM and the PGD method.}
\label{fig:Black-box attack}
\end{figure*}

\begin{table}[t!]
  \caption{\label{tab:cmresult} {Anti-spoofing performance of the three countermeasure models, LCNN-big, LCNN-small and SENet12.}}
  \centering
  \vspace{+0.5cm} 
\resizebox{\columnwidth}{!}{
\begin{tabular}{|c|c|c|c|c|}
\hline  
\multirow{2}{4em}{Model} &\multicolumn{2}{|c|}{Dev} &\multicolumn{2}{|c|}{Eval} \\ 
                  \cline{2-5}
            & t-$\text{DCF}_{norm}^{min}$  & EER(\%) & t-$\text{DCF}_{norm}^{min}$  & EER(\%) \\
                  \hline
LCNN-big          &0.0010       &0.047	        &0.1052	       &3.875          \\
LCNN-small        &0	        &0.002	        &0.1577	       &6.226          \\
SENet12           &0	        &0	            &0.1737	       &6.077          \\
\hline
\end{tabular}}
\end{table}

\begin{table}[t!]
  \caption{\label{tab:fgsmPGDwb} {White-box attack performance of the FGSM and the PGD method.}}
  \centering
  \vspace{+0.5cm}
\resizebox{\columnwidth}{!}{
\begin{tabular}{|c|c|c|c|c|c|c|}
\hline
\multirow{2}{4em}{EER(\%)} &\multicolumn{2}{|c|}{$\epsilon=0.1$} &\multicolumn{2}{|c|}{$\epsilon=1$} &\multicolumn{2}{|c|}{$\epsilon=5$}         \\ 
            \cline{2-7}
            & FGSM  & PGD        & FGSM   & PGD         & FGSM   & PGD          \\
            \hline
LCNN-big    &4.691  &\bf{6.256}  &36.504  &\bf{54.382}  &48.457  &\bf{93.119}   \\
LCNN-small  &7.613  &\bf{17.419} &34.670  &\bf{73.649}  &48.375  &\bf{89.845}   \\
SENet12     &7.737  &\bf{13.896} &24.936  &\bf{62.681}  &51.626  &\bf{87.220}   \\
\hline
\end{tabular}
}
\end{table}


\subsection{Adversarial attack results}
The subjective XAB listening test in subsection~\ref{subsec:XAB} results in average classification accuracy of 48.4\%, which confirms the validity of the adversarial audio attacks. Spectrograms of original audio, adversarial audio and adversarial perturbation of the utterance LA\_E\_1001227 is shown in Fig.~\ref{fig:spec}, where the attack is conducted to the LCNN-big model using the PGD method, with $\epsilon=5$, number of iterations $=$ 10 and number of random restarts $=$ 5. We can see that the adversarial perturbed spectrogram is almost visually indistinguishable from that of the original audio signal. 
If we set the EER point of the evaluation set as the operating point, the utterance LA\_E\_1001227 is wrongly classified as "bona fide" in this setting.
In the following subsections, we use the EER as metric to evaluate the performance of adversarial attacks by the means of the FGSM or the PGD method, under both white-box and black-box scenarios.

\subsubsection{White-box attacks}

The white-box attack performance of the FGSM and the PGD method using different $\epsilon's$ is shown in Table~\ref{tab:fgsmPGDwb}. The EERs of all three countermeasure models increase as $\epsilon$ grows. PGD attacks attain larger EERs than FGSM attacks in all three countermeasure models and all the settings of $\epsilon$ under the white-box attack scenario. The EERs of all three models under the FGSM attacks reach near 50\% when $\epsilon=5$. As for PGD, the EERs are greater than 50\% when $\epsilon = 1$, and are greater than 85\% when $\epsilon=5$, which will result in reversed classification decision if the operating point is pre-defined by the EER point on the evaluation set.
We can conclude from the white-box attack results that the reliability of all three countermeasure models are challenged and broken down by FGSM or PGD attacks under the scenario of white-box attack. The PGD method is more effective than the FGSM. Research on more advanced countermeasure models should be done to keep pace with today's white-box adversarial attacks.

\subsubsection{Black-box attacks}
In this part, we study adversarial attacks of the FGSM and the PGD method from the perspective of black-box scenario. As shown in Fig~\ref{fig:Black-box attack}, the black-box attacks achieve a resounding success in all the mutual attack settings: LCNN-big with LCNN-small, LCNN-big with SENet12 and LCNN-small with SENet12. In a large fraction of attacking scenarios, the attack method of PGD attains larger EER than the FGSM method, leading to that the PGD method tends to generate more powerful adversarial examples. According to Fig. \ref{fig:Black-box attack.a} and \ref{fig:Black-box attack.b}, adversarial examples generated by LCNN-big with rather more parameters are more powerful as they can attack smaller models, LCNN-small and SENet12, with larger EERs, while adversarial examples generated by LCNN-small and SENet12 fail to attack LCNN-big with such large EERs. So in our experimental setup, we can safely conclude adversarial examples from small models are outperformed by large models in terms of the performance of attack for both the FGSM and PGD methods. 
According to the red dotted line with triangle marker and red dash-dot line with circle marker in Fig. \ref{fig:Black-box attack.a} and Fig. \ref{fig:Black-box attack.b}, the adversarial examples generated by LCNN-big attain greater EER when attack LCNN-small rather than RESNet12, resulting in a conclusion that adversarial attacks are much easier realized under similar model structure.
According to Fig~\ref{fig:Black-box attack.c}, the attack efficacy of adversarial examples generated by LCNN-small outperforms the adversarial examples from SENet12. 


\section{Conclusions}
In this paper, we have investigated the vulnerability of spoofing countermeasures for ASV under both white-box and black-box adversarial attacks using the FGSM and PGD method. We have also compared performance of black-box attacks across spoofing countermeasure models with different network architectures and different amount of parameters. We implement three countermeasure models, i.e., LCNN-big, LCNN-small and SENet12, and conduct adversarial attacks on them. 
The experimental results show that all three models are subject to FGSM and PGD attacks under the scenario of white-box attack. The more dangerous black-box attacks are also proved to be effective by the experimental results. 
For the future work, we would like to adopt adversarial training methods to improve the robustness of countermeasure models and make them less vulnerable to adversarial attacks.

\section{Acknowledgements}
Songxiang Liu was supported by the General Research Fund from the Research Grants Council of Hong Kong SAR Government (Project No. 14208718). Haibin Wu and Hung-yi Lee were supported by the Ministry of Science and Technology of Taiwan (Project No. 108-2636-E-002-001).



\bibliographystyle{IEEEbib}
\bibliography{strings,refs}

\end{document}